\documentclass[twocolumn,tighten]{aastex63}
\pdfoutput=1 
\usepackage{amsmath,amstext}
\usepackage{apjfonts}
\usepackage{afterpage}
\usepackage{xcolor}
\usepackage{threeparttable}
\usepackage{graphicx}
\usepackage{epsfig}
\usepackage{booktabs}
\usepackage{multirow}
\usepackage{float}
\usepackage{mathrsfs}
\DeclareUnicodeCharacter{1E11}{\k{a}}
\setwatermarkfontsize{100px}

\begin{document}

\author{Tong Zhao}
\affiliation{School of Physics, Georgia Institute of Technology, 837 State St NW, Atlanta, GA 30332, USA}
\affiliation{Departments of Astronomy and Physics, University of Arizona, 933 N. Cherry Ave., Tucson, AZ 85721, USA}

\author[0000-0003-1035-3240]{Dimitrios Psaltis}
\affiliation{School of Physics, Georgia Institute of Technology, 837 State St NW, Atlanta, GA 30332, USA}

\author[0000-0003-4413-1523]{Feryal \"{O}zel}
\affiliation{School of Physics, Georgia Institute of Technology, 837 State St NW, Atlanta, GA 30332, USA}

\title{A Counterintuitive Correlation Between Neutron Star Radii Inferred from Pulse Modeling and Surface Emission Beaming}

\begin{abstract}
Thermal X-ray emission from rotation-powered millisecond pulsars, shaped by gravitational lensing and the beaming of the surface radiation, provides critical insights into neutron star properties. This approach has been the focus of observations with the NICER mission. Using a semi-analytic model to calculate pulse profiles, we investigate the effects of adopting incorrect beaming models on the inferred compactness of neutron stars. We demonstrate that assuming a more centrally peaked beaming pattern when fitting data from a more isotropic emitter leads to an underestimation of compactness in the case of two antipodal polar caps. We present a detailed analysis of this counterintuitive result, offering both qualitative insights and quantitative estimates. If the atmospheric heating in the millisecond pulsars observed with NICER is shallow, the inferred radii for these sources could be significantly overestimated, with important implications for neutron star structure and equation-of-state constraints.
\end{abstract}


\section{Introduction}

The thermal X-ray emission observed from isolated rotation-powered pulsars provides unique ways of measuring the properties of neutron stars and constraining the physics of their interiors~\citep[see][for a review]{Ozelreview2013}. Young pulsars with ages $\lesssim 10^6$~years emit the latent heat from their formation and are detected in soft X-rays \citep[see, e.g.,][]{Page2006}. In rotation-powered millisecond pulsars, which are significantly older, the observed thermal emission from their surfaces is thought to arise from bombardment of magnetospheric charges onto their polar caps~\citep{Ruderman1975, Arons1981}. 

In principle, the localized thermal emission from the polar cap of a spinning neutron star generates a pulsating flux of significant amplitude to an observer at infinity. However, general relativistic light bending suppresses the amplitude of this pulsation by an amount that depends on the mass-to-radius ratio (i.e., the compactness) of the star~\citep{spot1983}. The Neutron star Interior Composition ExploreR \citep[NICER;][]{NICER2016} was designed to exploit this effect and measure the radii of several nearby rotation-powered millisecond pulsars by fitting model pulse profiles to the periodic modulations of their X-ray fluxes~\citep{NICER1,NICER2}. To date, constraints on the properties of three sources, PSR~J0030+0451, PSR~J0740+6620, and PSR~J0437$-$4715, have been reported based on NICER observations~\citep{Riley2019, Riley2021, Miller2019, Miller2021,Vinciguerra2024,Choudhury2024}. 

It has been understood for many decades that the degree of beaming of the radiation emerging from the neutron star surface competes with gravitational lensing to alter the observed amplitudes of pulsations from a spinning neutron star~\citep{Riffert1988}: a more isotropic beaming results in a smaller pulsation amplitude and vice versa. In a companion article~\citep{Zhao2024}, we demonstrated that the presence of significant beaming is, in fact, necessary in lifting degeneracies between various model parameters. These necessitate special care in utilizing appropriate models of the surface emission from the stellar surface. 

The beaming of radiation emerging from a patch on a deep-heated neutron star atmosphere is determined entirely by the local gravitational acceleration and the effective temperature and is peaked towards the surface normal~\citep[see, e.g.,][]{Zavlin1996,Rajagopal1996,Haakonsen2012}\footnote{The beaming also depends on the composition of the atmopshere; in the case of recycled millisecond pulsars, the atmosphere is expected to consist purely of hydrogen, unless its companion star was depleted of it.}. However, for an atmosphere bombarded by magnetospheric charges, the beaming depends also very strongly on the energy spectra of the bombarding charges. Using a detailed physical model of heat deposition, \citet{Baubock} showed that bombarding charges with shallow energy spectra deposit most of their energies well below the photosphere and lead to radiation that is beamed towards the normal to the surface. In contrast, charges with steep energy spectra deposit most of their energies into the photosphere, leading to more isotropic beaming patterns~\citep[see, also,][]{Salmi2020}.

The impact of the assumed beaming pattern on the inference of neutron star properties via pulse-profile modeling has been explored in earlier work, both in the context of the NICER sources~\citep{Salmi2023} as well as in the context of burst oscillations, which are believed to be caused by localized thermonuclear burning on the surfaces of accreting neutron stars~\citep{Xburst1, burst2}. All these studies showed that the beaming function of the emerging radiation has to be known to within $\sim 5$\% in order for it not to introduce significant biases or uncertainties in the results\footnote{Note that the externally-bombarded atmosphere model studied by~\citet{Salmi2023} has parameters such that the resulting beaming function is practically indistinguishable from that of a deep-heated atmosphere.}.

In this paper, we use the semi-analytic model for pulse profiles that we developed in~\citet{Zhao2024} to understand a counterintuitive bias that arises in the inference of the compactness of a neutron star using this method, when an incorrect model for the beaming function is used. To illustrate this bias, consider the situation in which we try to model the pulse profile from a neutron star in which the surface emission is nearly isotropic but we are utilizing, erroneously, a model with strong beaming. In principle, we would have expected to infer a larger compactness for the neutron star, such that gravitational lensing could counteract more efficiently the effect of the stronger assumed beaming. Instead, for the case of emission from two antipodal polar caps, we find that using a model with stronger beaming results in inferring a smaller compactness for the neutron star.
 
In \S2, we give a brief summary of the semi-analytic model and the synthetic data we use and then, in \S3, study the bias introduced by the use of an incorrect beaming functions. In \S4, we give both a qualitative explanation of this result, as well as a detailed quantitative exploration based on our semi-analytic model. In the last section, we summarize our results and discuss implications for the inference of neutron star radii from pulse modeling.

\begin{table}[t]
\caption{Beaming function parameters}
\begin{center}
\begin{tabular}{c c c c}
\toprule
$\delta$ & $a$ & $b$ & $c$ \\
\midrule
1 & 0.0500 & 0.3695 & -0.00976\\
2 & 0.0500 & 0.3106 & -0.00976\\
3 & 0.0500 & 0.0955 & -0.00976\\
\bottomrule
\end{tabular}
\end{center}
\label{table:fit}
\end{table}

\section{AN APPROXIMATE ANALYTIC MODEL FOR PULSE PROFILE MODELING}

In \citet{Zhao2024}, we developed an approximate semi-analytic model of pulsations arising from localized thermal emission on the surface of a neutron star of mass $M$ and radius $R$, such that its compactness is $u\equiv 2GM/Rc^2$. This semi-analytic approach is based on an approximate expression for light bending~\citep{Pout2} and the Schwarzschild+Doppler approximation for the spacetime~\citep{Pout1}. For the purposes of this model, we assumed that the flux we observe originates on two antipodal hot spots at the north and south poles of the dipolar magnetic field of the star, both of which share the same temperature.  Also, we consider the case for which the neutron star is slowly spinning (as is the case for the primary NICER targets) and the linear size of each polar caps is small compared to the neutron star radius.

A purely dipolar magnetic field and a uniform temperature distribution on the polar caps are approximations that are probably not formally valid for the NICER sources (see, e.g., \citealt{Will} for a study of the expected polar cap geometry and temperature profiles in the presence of quadrupolar components in the magnetic field). In this study, we also neglect Doppler effects. Although relaxing any of these assumptions may lead to additional degeneracies between model parameters, it will not affect the origin or magnitude of the counterintuitive correlation that we are exploring here.

We assume that the emission at photon energy $E$ is locally described by the blackbody function $I_b(E,T)$ at temperature $T$ and that its angular dependence is quantified by the beaming function 
\begin{equation}
I(E,\alpha)=I_b(E,T)\frac{1+h(E,T)\cos\alpha}{1+(2/3)h(E,T)}\;.
\end{equation}
Here, $I(E,\alpha)$ is the intensity of emerging radiation, $\alpha$ is the angle with respect to the surface normal at the center of the spot, and $h(E,T)$ is the beaming factor. The factor $1+2/3h(E,T)$ in the denominator is used in this expression to preserve the total flux emerging from the surface.

In~\citet{Zhao2024}, we devised a quadratic functional form of the beaming factor $h$ that approximates the results of the bombarded atmospheric models calculated in~\cite{Salmi2020}, i.e.,
\begin{equation}
h(E,T)=a+b\left(\frac{E}{kT}\right)+c\left(\frac{E}{kT}\right)^2\;.
\label{eq:beaming_energy}
\end{equation}
The coefficients $a$, $b$ and $c$ depend on the power law index $\delta$ of the energy distribution of particles in the return current and are shown for completness in Table~\ref{table:fit}. 

\begin{table}[t]
\caption{Weakly Degenerate Parameters}
\begin{center}
\begin{tabular}{c c c c}
\toprule
parameters & description & units & range \\
\midrule
$q$ & $u+(1-u)\cos\theta \cos\zeta$ & dimensionless & $0<q<1$ \\
$s$ & $u-(1-u)\cos\theta \cos\zeta$ & dimensionless & $-1<s<0.5$\\
$p$ & $(1-u)\sin\theta \sin\zeta$ & dimensionless & $0<p<1$ \\ 
$T_\infty$ & $T\sqrt{1-u}$ & keV & $>0$ \\
$A$ & $dS/D^2$ & dimensionless & $>0$\\
\bottomrule
\end{tabular}
\end{center}
\label{table:1}
\end{table}

Under these assumptions, the flux observed at infinity is determined by 7 parameters: the mass $M$ and radius $R$ of the neutron star, the inclination angle $\theta$ of the observer with respect to the stellar spin axis, the colatitude $\zeta$ of the north polar cap, the effective temperature $T$ of the surface emission, the surface area $dS$ of each polar cap, and the distance $D$ to the observer. In \citet{Zhao2024}, we introduced five combinations of model parameters ($p$, $q$, $s$, $T_\infty$ and an overall normalization factor $A$; see Table~\ref{table:1}), which are only weakly degenerate in their influence on the model predictions. Use of these parameters not only allows us to sample efficiently the posteriors when fitting NICER data but have also assisted in identifying a number of previous unexpected degeneracies between them. In this paper, we will again use these parameters to understand the unexpected sign of the bias introduced when using the incorrect beaming function to describe the emerging radiation. 

Written in terms of these weakly-degenerate parameters, the total flux observed at infinity at photon energy $E$ and rotational phase angle $\phi$ is
\begin{equation}
F(E,\phi)=F_1(E,\phi)+F_2(E,\phi),
\end{equation}
where $F_1$ and $F_2$ are the fluxes from the first and second spot, respectively. The flux from the first spot is given by
\begin{equation}
F_1(E,\phi)=
\begin{cases}
    {\cal F}_1(E,\phi)\;, &\text{if } u+(1-u)\cos\psi>0\\
    0\;, & \text{otherwise}
    \label{eq:F_1}
\end{cases}
\end{equation}
where
\begin{equation}
\cos\psi \equiv\cos\theta \cos\zeta +\sin\theta \sin\zeta \cos\phi
\label{eq:psi}
\end{equation} 
is the cosine of the angle between the photon momentum and the normal to the stellar surface at the center of the spot,
\begin{equation}
{\cal F}_1(E,\phi)=q(1+\bar{h}q)\bar{F}(E)\left[1+r_1(E)\cos\phi+r_2(E)\cos\phi^2\right]
\label{eq:Fbar}
\end{equation}
\begin{equation}
\bar{F}(E)=\frac{A(1-u)^{3/2}}{1+(2/3)\bar{h}} I^\prime _b(E^\prime ,T),
\end{equation}
\begin{equation}
r_1(E)=\frac{p}{q}\left(1+\frac{\bar{h}q}{1+\bar{h}q}\right)\;,
\label{eq:r1}
\end{equation}
\begin{equation}
r_2(E)=\left(\frac{p}{q}\right)^2\frac{\bar{h}q}{1+\bar{h}q}\;,
\label{eq:r2}
\end{equation}
and the beaming factor $h$ is evaluated at
\begin{equation}
    \bar{h}=h(E/\sqrt{1-u},T).
\end{equation}
Similarly, the flux of the second spot is
\begin{equation}
F_2(E,\phi)=
\begin{cases}
    {\cal F}_2(E,\phi)\;, &\text{if } u-(1-u)\cos\psi>0\\
    0\;, & \text{otherwise}
    \label{eq:F_2}
\end{cases}
\end{equation}
where
\begin{equation}
{\cal F}_2(E,\phi)=s(1+\bar{h}s)\bar{F}(E)\left[1-t_1(E)\cos\phi+t_2(E)\cos\phi^2\right]
\end{equation}
\begin{equation}
t_1(E)=\frac{p}{s}\left(1+\frac{\bar{h}s}{1+\bar{h}s}\right),
\label{eq:t1}
\end{equation}
\begin{equation}
t_2(E)=\left(\frac{p}{s}\right)^2\frac{\bar{h}s}{1+\bar{h}s}\;.
\label{eq:t2}
\end{equation}

Here, primed quantities are measured on the neutron star surface. As equations~(\ref{eq:F_1}) and (\ref{eq:F_2}) show, the total flux of the antipodal spot model depends on the visibility of each spot. In order to reduce redundancy in the model, we define the first spot as the one that is facing the observer at zero rotational phase angle and assume, without loss of generality, that both the observer and the first spot are in the north hemisphere. Depending on the two orientation angles $\theta$ and $\zeta$, the first spot may disappear at a phase angle $\phi_1$ and become visible again at the symmetric phase angle $2\pi-\phi_1$. The second spot will not be visible at zero phase angle but will appear between phase angles $\phi_2$ and $2\pi-\phi_2$. According to the approximate light bending formula, these angles are given by 
\begin{equation}
    \cos\phi_1=\max\left[\frac{-q}{p},-1\right]
    \label{eq:phi1}
\end{equation}
and
\begin{equation}
\cos\phi_2=\max\left[\frac{s}{p},-1\right]\;.
\label{eq:phi2}
\end{equation}

Based on the values of the orientation angles $\theta$ and $\zeta$, we can define four classes of visibility~\citep[see also][]{Pout1}. In Class~I, the second spot is invisible during the entire spin period. In Class~II, the first spot is always visible while the second spot appears at $\phi=\phi_2$ and disappears at $\phi=2\pi-\phi_2$.
In Class~III, both of the two spots are always visible during the entire spin period. Finally, in Class~IV, the first spot becomes invisible at $\phi=\phi_1$ and reappears again at $\phi=2\pi-\phi_1$. 

Considering the various regions of spot visibility, the total flux $F_{total}$ can be rewritten as a piece-wise function
\begin{equation}
F_{total}(E,\phi)=
     \begin{cases}
       {\cal F}_1(E,\phi), &\text{if } \phi<\phi_2 \\&\text{or } \phi>(2\pi-\phi_2)\\
       {\cal F}_1(E,\phi)+{\cal F}_2(E,\phi),  &\text{if } \phi_2\le\phi\le\phi_1 \\&\text{or } (2\pi-\phi_1)\le\phi\le(2\pi-\phi_2)\\
       {\cal F}_2(E,\phi),  &\text{if } \phi_1<\phi<(2\pi-\phi_1).
     \end{cases}
\end{equation}

\subsection{Synthetic Data Generation and Bayesian Model Fitting}

In order to generate synthetic data, we consider the effective area and response matrix of NICER\footnote{Specifically, we use \texttt{nixtiaveonaxis20170601v002.arf} and  \texttt{nixitref20170601v001.rmf}}, as well as the effects of interstellar extinction using the model of \citet{Robert1983} for a hydrogen column density of $N_H=1.5\times 10^{20} cm^{-2}$, which is close to the fit results for PSR~J0740+6620 \citep{Miller2021,Riley2021}. 

In~\cite{Zhao2024}, we show that, if the peak of the X-ray spectrum of a source falls outside the energy range that is observed with NICER, the quality of model parameter inference is degraded. To reduce this effect, we assume in this paper that the effective temperature of the polar caps is $T=0.3$~keV and also consider a fiducial neutron star with compactness $u=0.35$.

With these considerations, we generated various sets of synthetic data by drawing, for each energy and time bin, values from a Gaussian distribution centered at the predicted value and with a standard deviation equal to $\sqrt{N}$, where $N$ is the number of photons. In order to minimize systematics related to the particular realization of noise in the data, we rescaled the overall normalization factor $A$ of the model such that the total number of photons in each configuration is equal to 3100000, which is ten times larger the total number of photons collected by NICER for PSR~J0740$+$6620 \citep[see][]{Miller2021,Riley2021}.

Finally, we use a Metropolis-Hastings Markov-Chain Monte Carlo (MCMC) algorithm with steps drawn from Gaussian distributions and no tempering, in order to perform Bayesian inference of the model parameters. For the priors over the model parameters, we use flat-top distributions with limits shown in Table~\ref{table:1}.

\begin{figure}[t]
	\centering
	\includegraphics[width=1\linewidth, keepaspectratio]{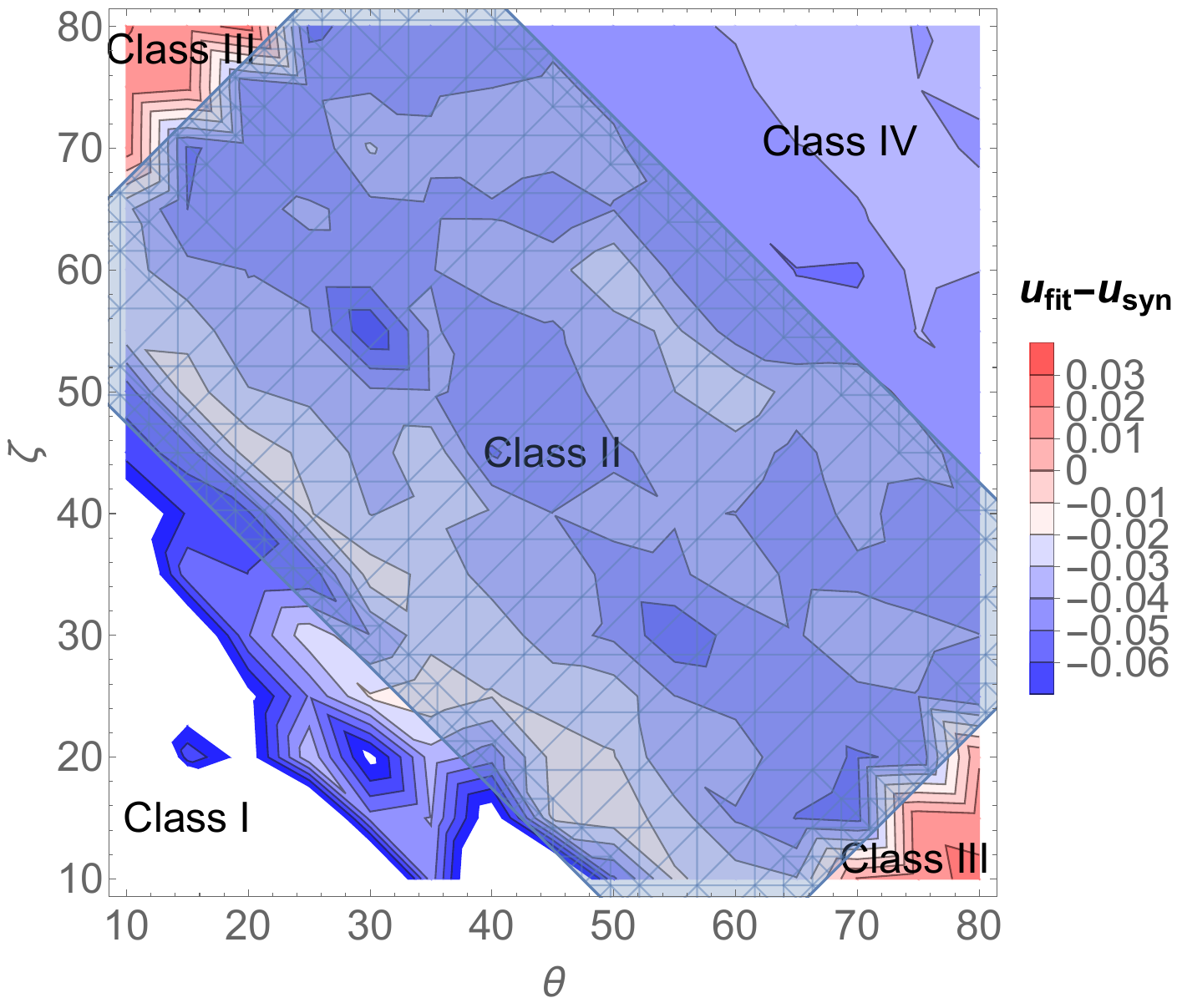}
	\caption{\label{fig:bias} Contours of the difference $u_{\rm fit}-u_{\rm syn}$ between the fitted compactness of a neutron star $u_{\rm fit}$  and the ground truth $u_{\rm syn}=0.35$ used in synthetic data, for difference values of the inclination of the observer $\theta$ and the colatitude of the polar caps $\zeta$. The synthetic data were generated with a beaming function that corresponds to shallow heating ($\delta=2$) but were fit with a model of a deep-heated atmosphere ($\delta=1$). Surprisingly, throughout the region of the parameter space that corresponds to visibility Class~II and IV, the best-fit value for the compactness is smaller than the ground truth.}
\end{figure}

\section{A counterintuitive consequence of utilizing an incorrect beaming function}\label{sec:bias}

In current analyses of NICER observations, the spectrum and beaming of the radiation emerging from the polar caps have been assumed to be those of a deep-heated atmosphere in radiative equilibrium. In such a deep-heating model, the magnetospheric charges are assumed to deposit their energy at layers much deeper than the photosphere causing the angular dependence of the emerging radiation (the beaming) to be peaked towards the surface normal. However, as discussed in the introduction, the thermal structure of a bombarded atmosphere depends strongly on the energy spectrum of the charges in the return current, which is not well understood. If the energy spectrum of the charges is steep, then the majority of their energy is deposited in the shallow photospheric layers, causing a significantly flatter beaming pattern. This is directly reflected in the dependence of the beaming parameter $b$ on the power-law index of the energy spectrum of the charges, as shown in Table~\ref{table:fit}.

\begin{figure}[t]
	\centering
	\includegraphics[width=1\linewidth, keepaspectratio]{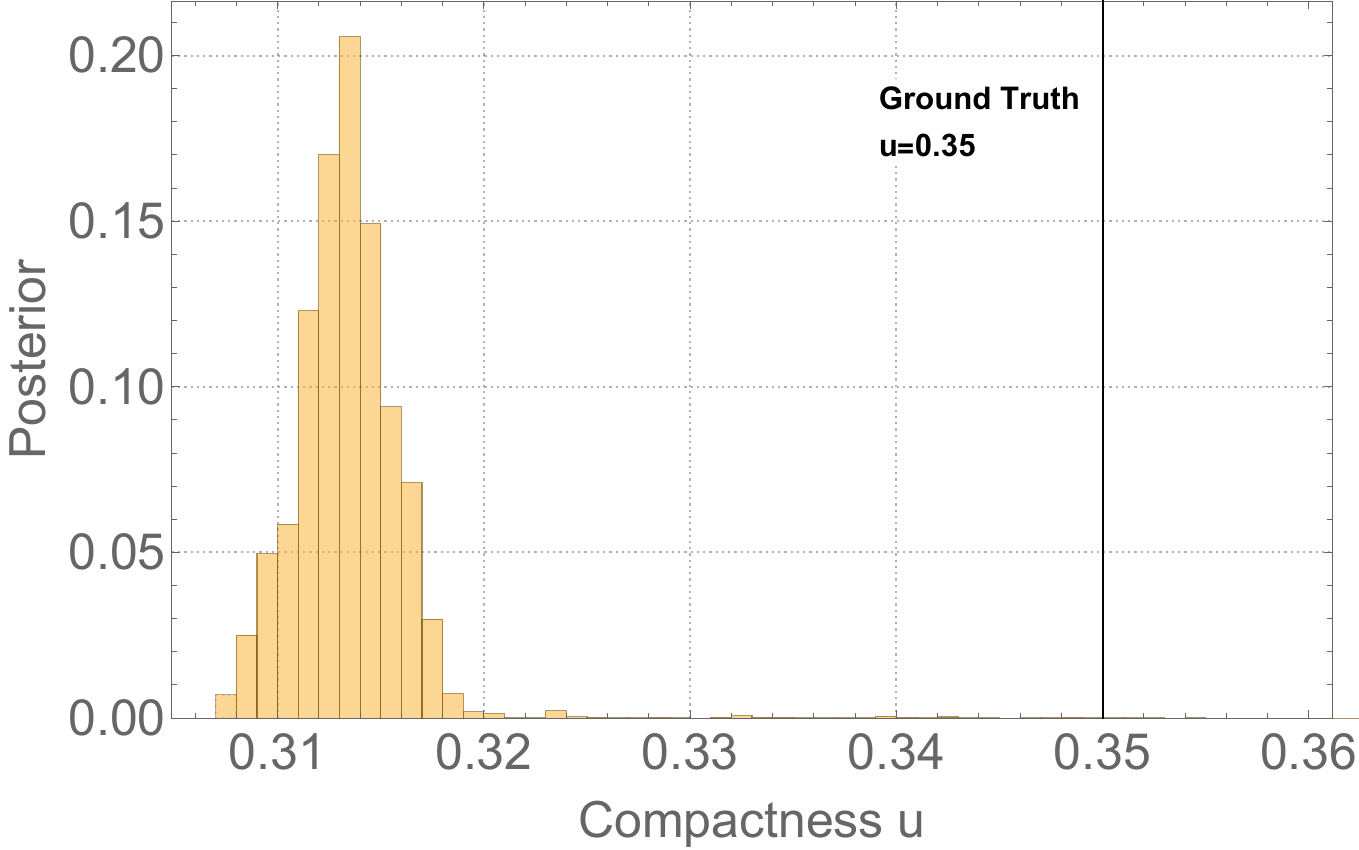}
	\caption{\label{fig:histogram} The posterior over the neutron star compactness $u$, for a configuration with an observer inclination of  $\theta=65^\circ$ and a colatitude of the polar caps of $\zeta=55^\circ$. The synthetic data were generated with a beaming function that corresponds to shallow heating ($\delta=2$) but were fit with a model of a deep-heated atmosphere ($\delta=1$). }
\end{figure}

There is an important question that arises naturally from this uncertainty: if the true beaming pattern of the emerging radiation is relatively isotropic (e.g., corresponding to a steep energy spectrum of charges with $\delta=2-3$) but the data are fit using a deep-heated atmosphere model, what would the effect on the inferred compactness $u$ of the neutron star be? The intuitive answer is that, utilizing a beaming pattern that is more centrally peaked than reality would cause the compactness of the neutron star to be increased. This way, the increased efficiency of gravitational self-lensing would reduce the amplitude of pulsations that would have been predicted by the model with more beaming. Surprisingly, we will show below that using a beaming function that is less  isotropic than the one assumed in generating synthetic data actually leads to an inferred compactness that is smaller than the ground truth value.

To demonstrate this, we generate synthetic data with a beaming function that corresponds to an energy spectrum of charges with $\delta=2$. Then, we fit the data with the same model but using a different beaming function, i.e., one that corresponds to $\delta=1$ that is similar to that of a deep-heated atmosphere. As discussed earlier, we assume an effective polar-cap temperature of $T=0.3$~keV and a compactness of $u=0.35$ for the synthetic data. We then perform Bayesian inference of the model parameters using 24 energy bins between 0.3~keV to 1.5~keV and 32 phase intervals per cycle. In order to ensure that our result is not specific to any particular observer inclination or spot colatitude, we generate and fit a large number of synthetic data sets with these two parameters ranging from $10^\circ$ and $80^\circ$ in increments of $5^\circ$.

\begin{figure*}[t]
	\centering
	\includegraphics[width=0.8\linewidth, keepaspectratio]{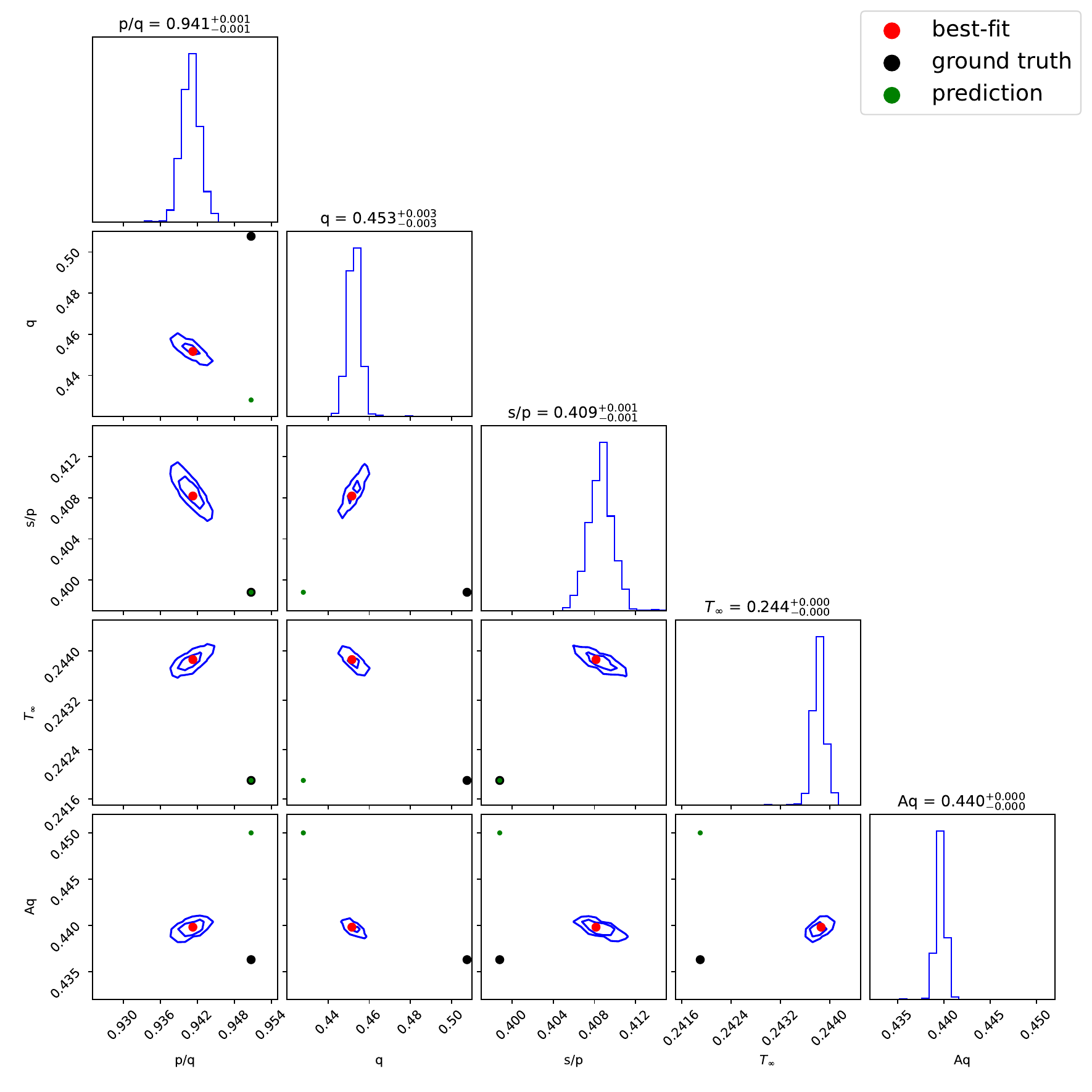}
	\caption{\label{fig:wrongbeamingcp} The posterior of fitting a model with deep heating ($\delta=1$) to synthetic data generated using a model with shallower heating ($\delta=2$). In this example, the observer inclination is $\theta=65^\circ$ and the colatitude of the polar caps is $\zeta=55^\circ$. Even though all projections of the posterior are very narrow, they are displaced from the ground-truth values shown as black circles. The model parameter $q$ displays the largest deviation from ground truth. The green circles show the bias estimated using the analytic results discussed in \S4.2.
 }
\end{figure*}

Figure~\ref{fig:bias} shows contours of the difference between the inferred compactness with the highest posterior $u_{\rm fit}$ and the one assumed for the synthetic data $u_{\rm syn}=0.35$, as a function of observer inclination $\theta$ and spot colatitude $\zeta$. On the same figure, we also outline the boundaries of the various visibility classes. Class~I corresponds to configurations in which only one of the two antipodal spots is visible and the amplitude of pulsations is very small. In the opposite extreme, Class~III corresponds to configurations in which both spots are always visible, the flux is nearly constant, and the small pulsations arise because of the presence of beaming. Because these extreme configurations are not relevant to the sources observed by NICER, we will not consider them here any further but instead focus on the more relevant Classes~II and IV. Figure~\ref{fig:bias} clearly shows that, in both Class~II and IV, the inferred value of the neutron star compactness is indeed smaller than the ground truth. 

We now focus on one particular set of values for the orientation angles, i.e., $\theta=65^\circ$ and $\zeta=55^\circ$, and compare directly the posterior over the compactness $u$ with the ground-truth value (see Fig.~\ref{fig:histogram}). The inferred compactness is $u_{\rm fit}\simeq 0.313$ which is about 10\% smaller than the ground truth $u_{\rm syn}=0.35$. At the peak of the photon spectrum (i.e., at a photon energy of $E=2 kT$), the fractional difference between the beaming factor $h$ of the model used for the synthetic data (i.e., the ground truth) and the one used for fitting them is $\sim 15$\%. Our results are, therefore, consistent with earlier studies in which the best-fit values of model parameters were significantly affected by the use of an incorrect beaming function, if the latter differ by more than 5\% from the ground truth~\citep{Salmi2023}.

In order to explore in more detail the origin of the sign of the bias, we show in Figure~\ref{fig:wrongbeamingcp} the corner plot of the posteriors over the weakly degenerate set of model parameters. Because of the small formal error in our synthetic data, the estimated values of the parameters have high precision but are indeed inaccurate. The largest fractional deviation occurs in the parameter $q$, which is also found to be 10\% smaller than the ground truth. As we will discuss below, this is the main driver behind the bias we found for the inferred compactness.

\section{Understanding the sign of the bias}

\subsection{A qualitative explanation}
\label{sec:qualitative}
In \S\ref{sec:bias}, we showed that using a model with large beaming to fit synthetic data generated with shallow beaming results in a best-fit value of the compactness that is smaller than the ground truth value. In order to understand qualitatively the sign of this bias, we will consider first the modulation of the flux that is generated by each of the polar caps. 

As shown in \citet{Zhao2024}, the modulation of the flux arises primarily by the evolution with rotational phase of the angle $\alpha$ with respect to the surface normal at which photons that reach the observer at infinity emerge from each spot, i.e.,
\begin{equation}
    F_i(E,\phi)\sim (dS\;\cos\alpha) \left[1+h(E,T)\cos\alpha\right]\;.
    \label{eq:F_i}
\end{equation}
In this expression, the term $(dS\; \cos\alpha)$ expresses the projected surface area of the polar cap towards the observer and the term in the square bracket quantifies the effect of the beaming. It is important to note here that the first of the two terms is achromatic, i.e., introduces the same modulation amplitude at all photon energies. In contrast, the term in the square bracket depends on photon energy through the beaming factor $h(E,T)$. Using the approximate analytic model, we can write the dependence of the beaming angle $\alpha$ on the rotational phase $\phi$ as
\begin{equation}
    \cos\alpha=\left\{
    \begin{array}{ll}
    q+p \cos\phi\;, & {\rm for~the~first~spot}\\
    s-p \cos\phi\;, & {\rm for~the~second~spot}.
    \end{array}
    \right.
\end{equation}
In other words, the parameters $q$ and $s$ measure the typical projections of the surface areas of the two polar caps to the distant observer, whereas the parameter $p$ measures their excursion with rotational phase (which is the same for both polar caps). Combining these two equations gives, for the first spot,
\begin{equation}
    F_1(E,\phi)\sim q (1+ h q)+p(1+2 h q )\cos\phi+...
\end{equation}
and for the second spot
\begin{equation}
  F_2(E,\phi)\sim s (1+ h s)-p(1+2 h s )\cos\phi+...
\end{equation}
This is the origin of expressions~(\ref{eq:r1}) and (\ref{eq:t1}) for the fractional amplitudes discussed earlier.

\begin{figure}[t]
	\centering
	\includegraphics[width=1.0\linewidth, keepaspectratio]{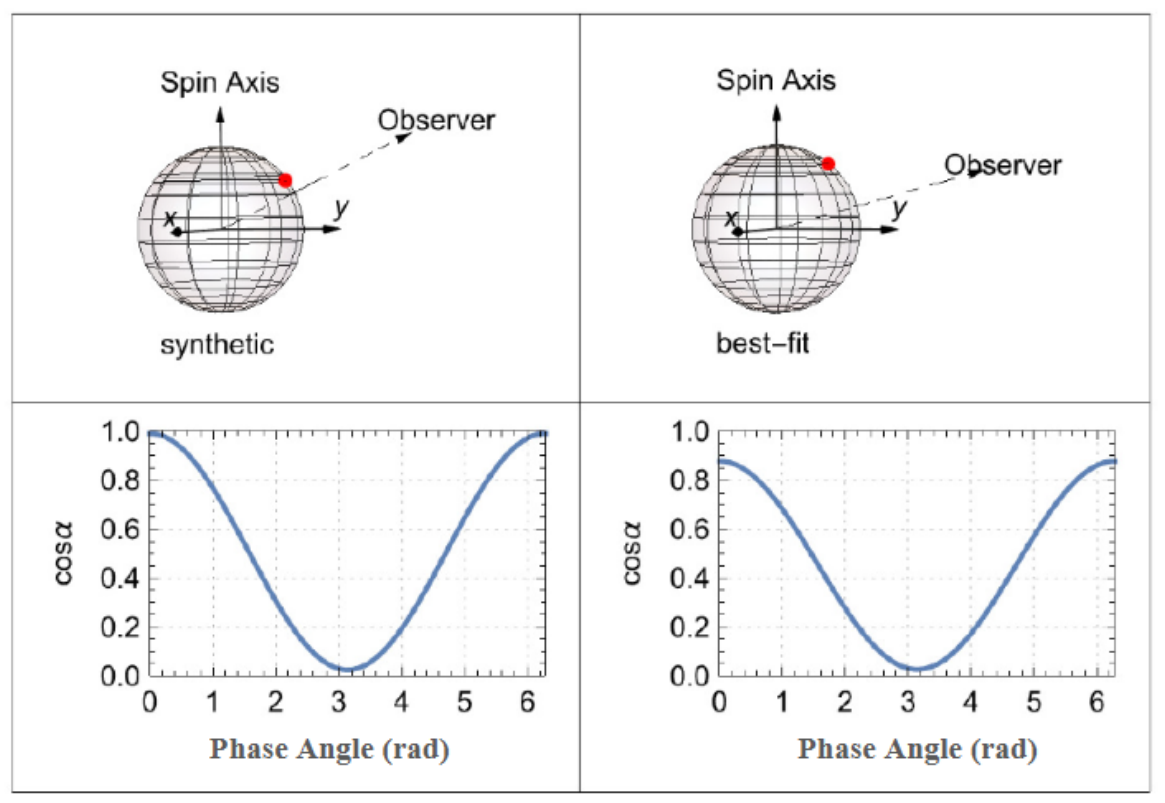}
	\caption{\label{fig:geometry} The geometry and the emission angle of the first spot as a function of the neutron star phase angle (in radians) for both the synthetic data and the best-fit result. In the best-fit result, the observer is looking at the spot at a larger angle with respect to its surface normal, which decreases the amplitude although with a smaller compactness.}
\end{figure}

Figure~\ref{fig:geometry} illustrates the geometry of the particular configuration explored in the example discussed around Fig.~\ref{fig:wrongbeamingcp} and shows the corresponding excursion of the emission angle $\alpha$ as a function of the rotational phase. The left panel corresponds to the configuration assumed in generating the synthetic data, whereas the right panel corresponds to the best-fit result. Comparing the two, we find that the angular distance between the observer and the polar cap is larger in the best-fit configuration than in the one used for the synthetic data. Because the observer is looking at the spot at a larger relative angle with respect to its surface normal, the mean value of $\cos\alpha$ (which is equal to $q$) decreases. However, the relative amplitude of its variation with rotational phase (which is equal to $p/q$) does not have a significant change because of the smaller compactness. It is this change in relative orientation that reduces the mean value of $\cos\alpha$ and causes the best-fit configuration to counter the effects of increased beaming.

But why is changing the relative orientation of the observer and the polar caps better than simply increasing the compactness? The reason is that changing the compactness alone primarily alters the achromatic effect of gravitational lensing, i.e., it changes the amplitudes of pulsations at all photon energies by a similar amount. In general, however, the beaming factor $h(E,T)$ is a function of photon energy and that introduces pulsation amplitudes that depend strongly on photon energy. In order for the best-fit model to generate the same pulsations as the synthetic data {\em at all photon energies\/}, the term in the square brackets in equation~(\ref{eq:F_i}), i.e., the one that is not achromatic, will need to be the one changing in order to accommodate for the change in the beaming function. For the first spot, this means that the product $h q$ needs to remain the same between the ground-truth and best-fit configuration and, for the second spot, the product $h s$ needs to remain the same. In other words, if we attempt to fit the data with a model that has stronger beaming (i.e., larger factor $h$) than the ground truth, then the mean projections of both the polar cap surface areas $q$ and $s$ will need to be reduced. 

\begin{figure}[t]
	\centering
	\includegraphics[width=1\linewidth, keepaspectratio]{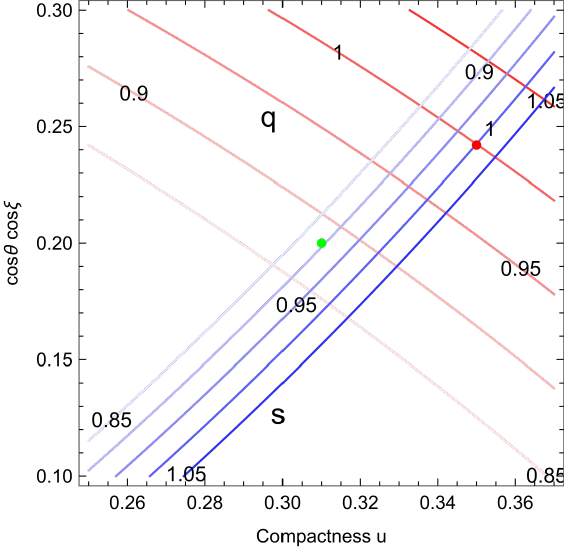}
	\caption{\label{fig:sandq} Contours of constant values of the model parameters $q$ and $s$ in the plane that contains the product $\cos\theta\cos\zeta$ and the neutron star compactness $u$. The values of $q$ and $s$ are normalized by their true values for the synthetic data in the example shown in Figure~\ref{fig:wrongbeamingcp}. The typical projection of the surface area of the first polar cap $q$ increases with increasing neutron star compactness $u$ and increasing value of the product $\cos\theta\cos\zeta$. On the other hand, the typical projection of the surface area of the second polar cap $s$ increases with increasing compactness $u$ but decreases with the product $\cos\theta\cos\zeta$. Accommodating a simultaneous decrease in both projection parameters $q$ and $s$ require a decrease in neutron star compactness. The red dot indicates the values of the parameters used in the synthetic data whereas the green dot indicates the values in the best-fit model with the incorrect beaming function.}
\end{figure}

Reducing the projection of the surface area of the first polar cap, i.e., reducing $q$, can be achieved either by reducing the compactness of the neutron star $u$, so gravitational lensing becomes less strong, or by decreasing the product $\cos\theta\cos\zeta$, so that the polar cap is viewed at a more grazing incidence (see red contours in Fig.~\ref{fig:sandq}). However, reducing also the projection of the surface area of the second polar cap, i.e., reducing also $s$, requires either to reduce the compactness $u$ or to {\em increase} the product $\cos\theta\cos\zeta$ (see blue contours in Fig.~\ref{fig:sandq}). The latter is a consequence of the antipodal geometry of the polar caps: the change that increases the viewing angle of the first polar cap will decrease, by necessity, the viewing angle of the second polar cap. The only combination that decreases simultaneously the projected surface area of both polar caps is one in which both the product $\cos\theta\cos\zeta$ and the compactness of the neutron star decrease. 

In summary, countering the effects of fitting the data with a model that has a stronger beaming function than the ground truth introduces a bias to both the relative orientation of the observer and the polar cap and to the neutron star compactness. In terms of the weakly-degenerate model parameters we introduced earlier, the bias affects primarily one of them, i.e., $q$ (because $s=s/p\times p/q\times q$, the parameter $s$ does not appear independently in the set shown in Table~\ref{table:1} but only in the ratio $p/s$, which is only marginally biased). This is not surprising given the fact that, as discussed in \citet{Zhao2024}, the parameter $q$ is completely unconstrained by data when the beaming of the emerging radiation is isotropic ($h=0$). It is the presence of substantial beaming that allows $q$ to be measured. Therefore, if the incorrect beaming function is used, the same parameter will be the one that is primarily affected.

.

\subsection{A quantitative measure of the bias}

In this section we provide a quantitative measure of the bias introduced by fitting pulse profile data with a model based on an incorrect beaming function. In particular, we will use the expansion of the flux in terms of its Fourier components
\begin{equation}
F(E,\phi)=A_0(E)+\sum_{n=1} A_n(E)\cos\left(n \phi\right)\;,
\end{equation}
with the explicit expression for the amplitudes $A_n$ given in \cite{Zhao2024}:
\begin{equation}
\begin{split}
A_0(E)=&\frac{q(1+\bar{h}q)\bar{F}(E)}{2\pi}\times\\\
&\qquad \left[(2+r_2)\phi_1+(2r_1+r_2\cos\phi_1)\sin\phi_1\right]\\
&+\frac{s(1+\bar{h}s)\bar{F}(E)}{2\pi}\times\\
 &\qquad\left[(2+t_2)(\pi-\phi_2)+(2t_1-t_2\cos\phi_2)\sin\phi_2\right]\;,
\end{split}
\label{eq:A0}
\end{equation}
\begin{equation}
\begin{split}
A_1(E)=&\frac{q(1+\bar{h}q)\bar{F}(E)}{\pi}\times\left[r_1\phi_1\right.\\
&\left.\qquad+\left(2+r_1\cos\phi_1+\frac{3}{2}r_2\right)\sin\phi_1+\frac{1}{6}r_2\sin(3\phi_1)\right]\\
&+\frac{s(1+\bar{h}s)\bar{F}(E)}{\pi}[t_1(\phi_2-\pi)\\
& \left.\qquad+\left(-2+t_1\cos\phi_2-\frac{3}{2}t_2\right)\sin\phi_2-\frac{1}{6}t_2\sin(3\phi_2)\right]\\
\end{split}
\label{eq:A1}
\end{equation}
and
\begin{equation}
\begin{split}
A_2(E)=&\frac{q(1+\bar{h}q)\bar{F}(E)}{\pi}\left[\frac{1}{2}r_2\phi_1+r_1\sin(\phi_1)\right.\\
&\qquad+\left(1+\frac{1}{2}r_2\right)\sin(2\phi_1)\\
&\qquad+\left.\frac{1}{3}r_1\sin(3\phi_1)+\frac{1}{8}r_2\sin(4\phi_1)\right]\\
&+\frac{s(1+\bar{h}s)\bar{F}(E)}{\pi}\left[\frac{1}{2}t_2(\phi_2-\pi)+t_2\sin(\phi_2)\right.\\
& \qquad +\left(-1+\frac{1}{2}t_2\right)\sin(2\phi_2)\\
&\qquad+\left.\frac{1}{3}t_1\sin(3\phi_2)+\frac{1}{8}t_2\sin(4\phi_2)\right]\;.
\end{split}
\label{eq:A2}
\end{equation}
Note that we have suppressed the expansion into sines, since we have neglected Doppler effects, which are the ones contributing to the sine expansion.

For ease of notation, we define $x\equiv p/q$ and $y\equiv s/p$ and consider only configurations in visibility Class~II for simplicity, for which $\phi_1=2\pi$. In this case, we can write the various Fourier amplitudes as linear functions of the beaming factor $\bar{h}(E,T)$ with five free parameters: $q$, $x$, $y$, $T_\infty$, and $A$. Since the only dependence of these amplitudes on photon energy comes through the common factor $\bar{F}(E)$ and the beaming factor $\bar{h}(E,T)$, we can write them schematically as
\begin{equation}
A_0=\bar{F}(C_1+C_2\; \bar{h}),
\end{equation}
\begin{equation}
A_1=\bar{F}(C_3+C_4\; \bar{h}),
\end{equation}
and
\begin{equation}
A_2=\bar{F}(C_5+C_6\; \bar{h}).
\end{equation}
Here the coefficients $C_1-C_6$ are all independent of $T_\infty$ and $A$. 

Using this notation, we can now write the fractional amplitude of the fundamental Fourier component as
\begin{equation}
\frac{A_1}{A_0}=\frac{C_3+C_4\; \bar{h}(E,T)}{C_1+C_2\; \bar{h}(E,T)}=\frac{(C_3/C_1)+(C_4/C_1)\bar{h}(E,T)}{1+(C_2/C_1)\bar{h}(E,T)}.
\end{equation}
with
\begin{equation}
\frac{C_3}{C_1}=\frac{x}{\pi}\frac{\phi_2-xy\sqrt{1-y^2}}{1+xy+\frac{x\sqrt{1-y^2}}{\pi}-\frac{xy\phi_2}{\pi}},
\label{eq:c3c1}
\end{equation}
\begin{equation}
\frac{C_4}{C_1}=\frac{qx}{\pi}\frac{2\pi-2\pi xy-\frac{3}{2}x\sqrt{1-y^2}+2xy\phi_2-\frac{1}{6}x\sin(3\phi_2)}{1+xy+\frac{x\sqrt{1-y^2}}{\pi}-\frac{xy\phi_2}{\pi}},
\label{eq:c4c1}
\end{equation}
and
\begin{equation}
\frac{C_2}{C_1}=q\frac{1+\frac{1}{2}x+\frac{3}{2\pi}x^2y\sqrt{1-y^2}+\frac{x^2}{2\pi}(1+2y^2)(\pi-\phi_2)}{1+xy+\frac{x\sqrt{1-y^2}}{\pi}-\frac{xy\phi_2}{\pi}}\;.
\label{eq:c2c1}
\end{equation}
Finally, we will use the subscript ``syn'' to denote the values of the various parameters for the model that was used in the generation of the synthetic data set and ``fit'' for the model with the incorrect beaming function that was used to fit the data.

In order for the incorrect model to reproduce the modulation amplitude $A_1/A_0$ of the synthetic data at all photon energies, the following relations need to be satisfied
\begin{eqnarray}
\left(\frac{C_3}{C_1}\right)_{\rm fit}&\approx&\left(\frac{C_3}{C_1}\right)_{\rm syn}\;,\label{eq:C3C1eq}\\
\left(\frac{C_4}{C_1}\right)_{\rm fit} \bar{h}(E,T)_{\rm fit}&\approx&
\left(\frac{C_4}{C_1}\right)_{\rm syn}\bar{h}(E,T)_{\rm syn}\;,\\
\left(\frac{C_2}{C_1}\right)_{\rm fit}\bar{h}(E,T)_{\rm fit}&\approx&
\left(\frac{C_2}{C_1}\right)_{\rm syn}\bar{h}(E,T)_{\rm syn}\;.\label{eq:C2C1eq}
\end{eqnarray}
The last two approximate equations cannot be formally satisfied at all photon energies for the functional form of the beaming function we are using. However, we can estimate the magnitude of the bias by evaluating them at the photon energy of the peak of the X-ray spectrum. For the example discussed in the previous section, $T_\infty=0.24$ keV, the number of photons for a black body spectrum peaks at $E\simeq 0.4$~keV and, at that photon energy, $\bar{h}_{syn}/\bar{h}_{\rm fit}\simeq 0.85$.

The system of equations~(\ref{eq:C3C1eq})-(\ref{eq:C2C1eq}) has the unique solution
\begin{equation}
    \left(x_{\rm fit},y_{\rm fit},q_{\rm fit}\right)=\left(x_{\rm syn}, y_{\rm syn}, q_{\rm syn} \frac{\bar{h}_{\rm syn}}{\bar{h}_{\rm fit}}\right)\;,
    \label{eq:analytic_bias}
\end{equation}
from which we conclude that bias is introduced primarily to the parameter $q$, as discussed earlier. 

We can now use the relations shown in Table~\ref{table:1} and solve for the physical model parameters $\theta$, $\zeta$, and $u$ to obtain
\begin{equation}
\cos(\theta+\zeta)=\frac{1-xy-2x}{2-(1+xy)q}q\;,
\end{equation}
\begin{equation}
\cos(\theta-\zeta)=\frac{1-xy+2x}{2-(1+xy)q}q\;,
\end{equation}
and
\begin{equation}
u=\frac{xy+1}{2}q\;.
\end{equation}
This last expression allows us to quantify the bias introduced by the use of the incorrect beaming function to the inference of the neutron star compactness, i.e.,
\begin{equation}
u_{\rm fit}\approx u_{syn} \left(\frac{\bar{h}_{syn}}{\bar{h}_{\rm fit}}\right)\;.
\end{equation}
In other words, fitting data with a model that has stronger beaming than the ground truth, i.e., $h_{\rm fit}>h_{\rm syn}$, leads to an inferred compactness for the neutron star that is smaller than the true value, $u_{\rm fit}<u_{\rm syn}$.

As a final check, we consider the second order Fourier component of the pulse profile for which
\begin{equation}
\frac{A_2}{A_0}=\frac{(C_5/C_1)+(C_6/C_1)\bar{h}(E,T)}{1+(C_2/C_1)\bar{h}(E,T)}\;,
\end{equation}
where
\begin{equation}
\frac{C_5}{C_1}=\frac{x}{\pi}\frac{(1-y^2)^\frac{3}{2}}{1+xy+\frac{x\sqrt{1-y^2}}{\pi}-\frac{xy\phi_2}{\pi}}\;,
\end{equation}
and
\begin{equation}
\frac{C_6}{C_1}=\frac{qx^2}{\pi}\frac{\pi+\frac{1}{6}y(5-2y^2)\sqrt{1-y^2}-\frac{1}{2}\phi_2}{1+xy+\frac{x\sqrt{1-y^2}}{\pi}-\frac{xy\phi_2}{\pi}}\;.
\end{equation}
A trivial substitution shows that the solution~(\ref{eq:analytic_bias}) that we obtained for the fractional amplitude of the fundamental Fourier component also satisfies these equations.

Figure~\ref{fig:wrongbeamingcp} shows with green circles the predicted ``best-fit'' values of the various model parameters, as estimated with the approximate expressions derived in this section. The small fractional differences between these predicted values and those obtained by the Markov-Chain Monte Carlo method can be attributed primarily to the approximation involved in solving the system of equations~(\ref{eq:C3C1eq})-(\ref{eq:C2C1eq}).

\section{Conclusions}

In this paper, we used an approximate semi-analytic model for pulse profiles from spinning neutron stars that we had devised earlier~\citep{Zhao2024} in order to explore biases in parameter estimation introduced by utilizing an incorrect beaming function for the radiation emerging from the neutron star surface.
We showed that utilizing a model for a deep heated atmosphere, which has strong beaming, to fit data from a shallow-heated neutron star, which has weaker beaming, leads to an underestimate of the compactness of the neutron star, contrary to naive expectations.

Using fitting of synthetic NICER data as well as semi-analytic estimates, we showed that the fractional bias in the neutron star compactness is approximately equal to the fractional error in the assumed beaming factor $h$ at the peak of the energy spectrum and, in particular, that $(\delta u/u)\simeq - (\delta h/h)$. This is consistent with earlier estimates concluding that the beaming function needs to be known to within $\sim 5$\% in order to have insignificant influence on the inference of neutron star properties.

The energy distribution of the magnetospheric charges in the return current that heats the neutron star surface and, therefore, dictates the beaming function, is largely unknown. However, the deep-heated atmosphere models used in current analyses of NICER data have the strongest possible beaming of all alternatives. If the atmospheric heating is instead shallower, it will generate more isotropic beaming than what has been currently assumed. This will imply that the true values for the compactness of the neutron stars observed with NICER might be larger than what has been inferred. For a neutron star of known mass, a larger compactness will also imply a small radius, potentially relieving some of the tension with prior measurements of neutron star radii~\citep[see, e.g.,][]{Raithel2021}.

\bibliographystyle{apj}
\bibliography{MS}

\end{document}